# FRAMEWORK FOR A LOW-COST INTRA-OPERATIVE IMAGE-GUIDED NEURONAVIGATOR INCLUDING BRAIN SHIFT COMPENSATION

M. Bucki, C. Lobos and Y. Payan

Laboratoire TIMC-IMAG, UMR CNRS-UJF 5525, La Tronche, France

*Abstract* — In this paper we present a methodology to address the problem of brain tissue deformation referred to as 'brain-shift'. This deformation occurs throughout a neurosurgery intervention and strongly alters the accuracy of the neuronavigation systems used to date in clinical routine which rely solely on pre-operative patient imaging to locate the surgical target, such as a tumour or a functional area. After a general description of the framework of our intra-operative image-guided system, we describe a procedure to generate patient specific finite element meshes of the brain and propose a biomechanical model which can take into account tissue deformations and surgical procedures that modify the brain structure, like tumour or tissue resection.

## I. INTRODUCTION

Accurate localization of the target is essential to reduce the morbidity during a brain tumor removal intervention. Image guided neurosurgery is nowadays facing an important issue for large skull openings, with intra-operative changes that remain largely unsolved. In that case, deformations of the brain tissues occur in the course of surgery because of physical (dura opening, gravity, loss of cerebrospinal fluid, actions of the neurosurgeon, etc) and physiological phenomena (swelling due to osmotic drugs, anaesthetics, etc), some of them still being not completely known. As a consequence of this brain-shift, the pre-operatively acquired images no longer correspond to reality and the pre-operative based neuro-navigation is therefore strongly compromised by intra-operative brain deformations. Some studies have tried to measure this intra-operative brain-shift. Hastreiter and colleagues [1] observed a great variability of the brain-shift ranging up to 24 mm for cortical displacement and exceeding 3 mm for the deep tumor margin; the authors claim for a non-correlation of the brain surface and the deeper structures. Nabavi and colleagues [2] state that the continuous dynamic brain shift process evolves differently in distinct brain regions, with a surface shift that occurs throughout surgery (and that the authors attribute to gravity) and with a subsurface shift that mainly occurs during resection (that the authors attribute to the collapse of the resection cavity and to the intra-parenchymal changes).

In order to face this problem, scientists have proposed to add to current image-guided neurosurgical systems a module to compensate brain deformations by updating the pre-operative images and planning according to intra-operative brain shape changes. The first proposed algorithms deformed the pre-operatively acquired images using image-based models. Different non-rigid registration methods were therefore provided to match intra-operative images (mainly MRI exams) with pre-operative ones [3-5]. More recently, biomechanical models of the brain tissues were proposed to constrain the image registration: the models are used to infer a volumetric deformation field from correspondences between contours [6-7] and/or surfaces [8] in the images to register. Arguing against the exorbitant cost of the intra-operative MRI imaging devices, some authors have proposed to couple the biomechanical model of the brain with *low-cost* readily available intra-operative data [9] such as laser-range scanner systems [10-11] or intra-operative ultrasound [12]. This proposal seems appealing from a very practical point of view, compared with the *high cost* intra-operative MRI device. However, it gives to the biomechanical model a crucial and very central position. This means that a strong modelling effort has to be carried out during the design of the brain biomechanical model as well as its validation against clinical data.

This paper aims at introducing the methodology we propose to tackle the brain-shift problem, with very specific clinical and technical constraints:

- Before surgery:
  (1) a planning is defined from an MRI exam, with localization of the skull opening and the pathway towards the targeted tumor;
  (2) a patient-specific Finite Element (FE) model of the brain tissues is automatically built from the segmented cortical tissues and ventricles, with an FE mesh more refined along the pathway "skull opening/targeted tumor" and coarser elsewhere;
  (3) FE pre-computations are carried out in order to allow deformations of the biomechanical brain model during surgery within an acceptable delay i.e. 1 to 5 minutes.
- During surgery:
  (4) after craniotomy and dura opening, an aspiration device, the 'rheology pipette' (Schiavone et al., 2007), is used to measure the stiffness of the healthy cortex as well as the one of the tumor if it is located on the cortex surface. These parameters are used by the FE

Marek Bucki is with the TIMC-IMAG Laboratory, Equipe GMCAO, UMR CNRS-UJF 5525, Faculté de Médecine de Grenoble, Pavillon Taillefer, F38706 La Tronche Cédex; phone: +33 4 56 52 00 04; fax: +33 4 56 52 00 55; mail: Marek.Bucki@imag.fr

Claudio Lobos and Yohan Payan are also with the TIMC-IMAG Laboratory, UMR CNRS-UJF 5525, La Tronche, France. Email: firstname.lastname@imag.fr

model to predict with better accuracy the behaviour of the tissues within the region of interest, near the tumor;

(5) the *low-cost* solution is chosen, which means the absence of any intra-operative MRI data;

(6) to compensate this absence, localized 2.5D ultrasound (US) images are acquired during surgery, in order to track intra-operatively the position of a set of anatomical landmarks;

(7) the biomechanical model uses this input as new boundary conditions to estimate the brain deformations and the tumor position;

(8) finally a classical navigation procedure is used to guide the surgeon towards the target.

In the remainder of the paper we describe in detail our approach which still needs to be confronted with clinical reality as developments are under way.

## II. SURGICAL PLANNING

### A. Acquisition of pre-operative data and planning

Before the intervention, an MRI image volume of the patient's brain is acquired. Several segmentation algorithms [15] are available to extract from this volume the data necessary to produce surface meshes of the different structures we want to model. The more relevant ones are: the surface of the brain, the tumor, the ventricles and white matter tracts.

During the pre-operative planning stage an Opening Skull Point (OSP) of the brain must be specified. This information is used to produce a mesh that is more refined in the pathway from the OSP to the tumor. This localized refinement allows higher accuracy in the region where the resection of tissues will take place.

### B. Patient specific mesh generation

We propose a meshing technique that involves a conjunction of several algorithms with variants that allow optimizations in time-producing as well as element quality and quantity, in order to obtain a suitable mesh for a real-time application. The overall mesh process can be summarized in the following five steps:

**Basic non-regular octree mesh.** We use the octree technique [16-17] with the following constraints: (a) the subdivision stop condition will depend on the number of points in the mesh (fig. 1-A) and (b) only the elements that lie in the Region of Interest (RoI) will be subdivided. With this two constraints we obtain a mesh that is refined in the RoI, i.e. the pathway 'OSP – tumor' and coarse elsewhere. An example of this is shown in fig. 2.

**1-irregular octree mesh.** A mesh is said to be "1-irregular" when each element has a maximum of one point inserted in each edge or face. In other words, we split the elements that lie outside the RoI until every edge and face has a maximum of one point inserted on it. An example of this is shown in fig. 1-B.

**Transition management between different density regions.** We manage the transitions from refined to coarse zones inserting different types of elements like: pyramids, tetrahedrons and prisms. Our implementation considers different patterns that allow us to identify and split each possible combination of points inserted on the edges and faces [21]. An example of this is shown in fig. 1-C.

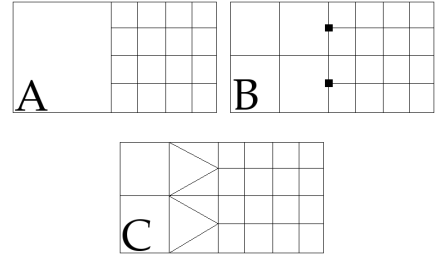

Fig. 1. Examples are shown in 2D even though the implementation considers the 3D cases. A) Basic octree with different zones of refinement, B) 1-irregular mesh where square points are inserted in edges of some elements and C) Mixed element mesh without points inserted on edges.

**Split all elements into tetrahedrons.** Every element of the mesh is subdivided into tetrahedrons because this type of element allows us to compute during the intervention, in an efficient way, the simulated MRI slices controlled by the biomechanical model deformation. This process doesn't increase the FE computation time because this bottleneck is related to the number of points and not the relation between themes.

**Projection of outside points onto the brain surface.** Finally the points of the elements that intersect the surface are projected to obtain a good representation of the brain.

We start with an octree approach because it easily allows the generation of meshes with refined and coarse zones giving us control over the quantity and quality of the elements. Another approach would be to start directly with a tetrahedron mesh but even though several techniques are available [18-20] none of them is as straightforward in the refinement by zones as the one presented here.

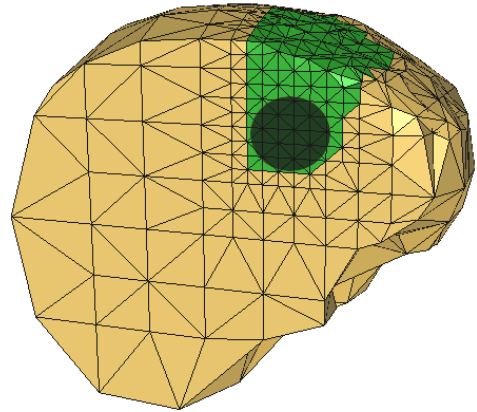

Fig. 2. Patient specific mesh and segmented tumor. The region of interest has a greater refinement level than the rest of the mesh. The tumor is represented by a circle in the most refined area.

Finally, bad quality tetrahedrons are removed keeping a consistent state of the final mesh. Note that bad quality elements can only be obtained after the projection onto the surface because before that step all tetrahedrons are built from good quality and regular elements therefore only superficial elements can be removed, avoiding this way, holes in the mesh. In other words, thanks to the initial octree approach we can guarantee a good element quality in each octant that has not been projected onto the surface as seen in fig. 2.

### C. FE model biomechanics and pre-computation

Miller has shown in his works [13] that brain tissue has a visco-elastic, non-linear behaviour which yields equations difficult to compute in real-time. A simpler model is thus necessary, especially if tissue resection needs to be modelled. Clatz et al. mention 7% [14] deformations in the case of Parkinson interventions with small apertures and we didn't find any results showing higher order deformation (10 or 15%). We thus chose, as a first approximation, a mechanical linear and small deformations model which will require validation against clinical data. This hypothesis allows us to model in an interactive way, surgical interactions such as cyst drainage and tissue resection.

The linear PDE are solved using the Finite Elements method. Within our FE mesh the nodes are labelled as 'pilot' i.e. nodes associated with anatomical structures and which displacements are updated using intra-operative US imaging; 'sliding' i.e. nodes in contact with the skull and which displacement is constrained along the skull tangent plane; 'free' i.e. nodes without any displacement constraint.

The linear mechanics PDE lead to a linear system $KU = F$ where K is the stiffness matrix, U the displacements at the nodes and F the forces applied on the nodes. The general solution of this system can be decomposed as a linear combination of elemental solutions computed for each 'pilot' node elemental displacement. These displacements are computed beforehand and stored.

Since our model is displacement-driven, the exact values of the Young modulus for grey matter, E(grey), and white matter, E(white), need not be known, only the ratio between these moduli matters. In the literature authors have proposed a wide range of values for white and grey matter. [22] propose E(grey)=180kPa and E(white)=18kPa; [23] propose E(grey)=E(white)=694Pa; [25] and [26] propose E(white)=E(grey)=3kPa. We chose to use a ratio between grey and white matter stiffness. The value chosen is an intermediate between those given by [22], where E(grey)/E(white)=10, and [24], where E(grey)/E(white)=2. Our model thus uses 'relative stiffness' coefficients: E(grey)=1 and E(white)=1/6.

## III. INTRA-OPERATIVE PROCESS

### A. Rigid registration and navigation of the craniotomy

Our system is implemented on a navigation station equipped with a passive optical localizer. Once the patient positioned in the OR, the pre-operative data are rigidly registered to the patient space. To this end, a series of points is acquired by sliding a tracked pointer on the patient's face. The resulting point cloud is registered onto the segmented skin within the pre-operative MRI volume and the skull opening is performed according to the planning initially defined, with the help of the neuronavigator.

### B. Tumor biomechanics

Cancer tissue stiffness has great variability depending on the pathology, and to our knowledge, no team has measured it. In the case where the tumor stiffness can be assessed using the rheology pipette, we take it into account the following way. First, we measure in vivo the stiffness of the healthy cortex, E(cortex). Then, the tumor stiffness value is measured in vivo, yielding E(tumor). The ratio E(tumor)/E(cortex) is the relative stiffness and we use it as an input to our model. The pre-computations described in II-C are updated using this value for the elements describing the tumor.

### C. 2.5D US intra-operative update

During the course of the intervention, whenever necessary, a set of localized US images is acquired by the surgeon. This 'pseudo-volume' contains anatomical structures that can be segmented either automatically or semi-automatically, such as ventricles. The pilot nodes are moved according to the positions of these structures and the global deformation is updated using these new positions as boundary conditions of the FE model. The location of the tumor is computed and the pre-operative MRI volume is deformed along with the segmented structures of interest such as white matter tracts near the resection area (see fig. 3).

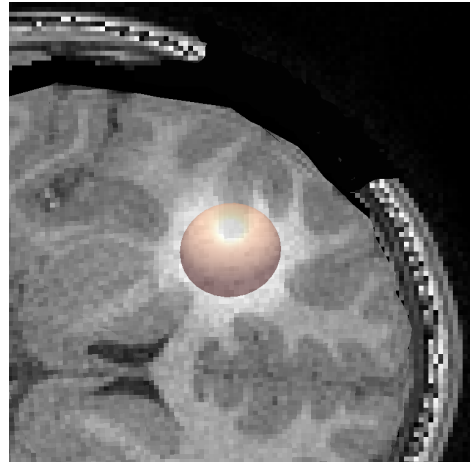

Fig. 3. Simulated MRI slice after dura opening (artificially removed on the image) and deeper structures settling due to gravity.

### D. Tissue resection modeling

Tissue resection or aspiration is modeled by local elements alteration.

In the case of tumors presenting a liquid cyst, the corresponding region can be easily segmented from a T2 weighted MRI scan. The cyst is modeled as an incompressible fluid by assigning a nearly incompressible Poisson parameter ($\upsilon$=0.49) and a relative Young module very low (E=0.01kPa) to its elements, as in [23]. When the aspiration is performed, the corresponding elements are removed from the mesh and the model deformation is updated in order to reflect the cavity collapse.

## IV. Results and Discussion

A mesh that takes into account patient-specific information on the pathway OSP-tumor has been presented. This approach allows high accuracy in the zone of study and creates fewer elements where lesser information is needed. The generation of a non-regular mesh is the most important contribution of this work in terms of meshing.

As for the intra-operative procedures, the linear FE model allows very fast deformation updates. For a 10200 tetrahedrons – 2700 nodes mesh, the deformation update based upon a set of about 100 'pilot' nodes displacements, along with MRI and segmented structures deformation, takes about 50 milliseconds on a PC P4 – 3GHz – 1Go RAM.

Using an optimized data structure to store and perform computations on sparse matrices and a specific node sorting to reduce computational complexity, the resection update delay ranges from a few seconds to, at most, 2 minutes, which is very acceptable in an OR context.

The definition and extraction of anatomical landmarks from localized 2D US images is a delicate task. The ventricles are an obvious choice due to their importance in the deformation process but other, less profound structures need also be defined in order to control in an accurate way our model.

The complexity of brain tissues (highly non linear mechanics, anisotropy and non-homogeneity) makes it difficult to achieve a realistic model with OR-compliant computational complexity and robustness. We think a robust neuronavigation system should mainly rely on dense anatomical reference extraction and use an interactive biomechanical model as 'smart interpolator' between this set of reliable landmarks. Localized 2D ultrasound, unlike intra-operative MRI, is very convenient for landmark tracking since it allows frequent updates during the intervention and presents the advantage of being a low-cost and widespread imaging device.